\documentclass[preprintnumbers]{revtex4}
\usepackage{amssymb}
\usepackage{amsmath}
\usepackage{graphicx}
\usepackage{dcolumn}
\usepackage{bm}
\usepackage{subfigure}
\usepackage{color}

\begin{document}

\title{Thermodynamics of phase transition in
Reissner-Nordstrom-de Sitter spacetime}
\author{Xin-Ping Li,$^{1,2}$ Yu-Bo Ma,$^{1,2}$  Yang Zhang,$^{1,2}$ Li-Chun Zhang,$^{2}$ Huai-Fan Li$^{1,2,*}$}
\address{$^1$Department of Physics, Shanxi Datong University, Datong 037009, China\\
$^2$Institute of Theoretical Physics, Shanxi Datong University, Datong, 037009, China }

\thanks{\emph{e-mail:huaifan.li@stu.xjtu.edu.cn}}

\begin{abstract}

The Reissner-Nordstrom-de Sitter (RN-dS) spacetime can be considered as a thermodynamic system. Its thermodynamic properties are discussed that the RN-dS spacetime has phase transitions and critical phenomena similar to that of the Van de Waals system or the charged AdS black hole. The continuous phase transition point of RN-dS spacetime depends on the position ratio of the black hole horizon and the cosmological horizon. We discuss the critical phenomenon of the continuous phase transition of RN-dS spacetime with Landau theory of continuous phase transition, that the critical exponent of spacetime is same as that of the Van de Waals system or the charged AdS black hole, which have universal physical meaning. We find that the order parameters are similar to those introduced in ferromagnetic systems. Our universe is an asymptotically dS spacetime,  thermodynamic characteristics of RN-dS spacetime will help us understand the evolution of spacetime and provide a theoretical basis to explore the physical mechanism of accelerated expansion of the universe.
\end{abstract}

\maketitle

\section{Introduction}

In recent years, the internal microstructure and the evolution of black holes has attracted great attention to the thermodynamic characteristics of AdS and dS spacetime. People define the cosmological constants of the n-dimensional AdS spacetime as the state parameters of the black hole thermodynamic system, i.e., pressure
\begin{equation}\label{1.1}
p=\frac{n (n-1)}{16 \pi l^{2}}, \quad \Lambda=-\frac{n (n-1)}{2 l^{2}},
\end{equation}
and the corresponding thermodynamic volume is
\begin{equation}\label{1.2}
V=\left(\frac{\partial M}{\partial P}\right)_{S,Q_{i},J_{k}}.
\end{equation}
The thermodynamic properties of AdS black holes have been studied extensively \cite{David Kubiznak12,Rong-Gen Cai13,Sharmila Gunasekaran12,Antonia M. Frassino14,Natacha Altamirano14,Natacha Altamirano13,Joy Das Bairagya20,X. H. Ge15,Rong-Gen Cai16,Jia-Lin Zhang15b,Ren Zhao13,Ren Zhao13b,Shao-Wen Wei09,S H Hendi17,A. Dehghani20,Rabin Banerjee20,M. Chabab2019,M. Chabab2016,Meng-Sen Ma17b,Dehyadegari2020,Daniela Mago2020,Sajadi2019,Hendi2019,Hendi2018,Eslam,Hendi2016a,Hendi2016b,Zou17,Jianfei15}. The study of the thermodynamic properties of black holes will not only deepen understanding of the nature of black holes, but also help to understand the microscopic particles inside black holes \cite{Shao-Wen19,Shao-Wen2020,Run Zhou08,Yu-Xiao20,Yan-Gang18,Yan-Gang2019,Xiong-Ying20,Xiong-Ying2019,Amin Dehyadegari17}. People first found that the charged AdS black hole has a characteristic similar to the Van de Waals phase transition when the independent variable $P-V$ was selected. It has been proved that for the charged AdS black hole, when the independent variable $Q^{2}-\Psi$ is selected, the black hole has a characteristic similar to the Van der Waals phase transition~\cite{Amin Dehyadegari17}

As we all know, in the early period of cosmological inflation, our universe is a quasi-de Sitter spacetime. If the cosmological constant corresponds to dark energy, our universe will evolve into a new de Sitter phase \cite{Cai02a}. In order to construct the entire evolution of the whole process and find the inner reason for the accelerated expansion of the universe, we should have a clear understanding of the classical, quantum properties and the thermodynamic properties of de Sitter spacetime, and have more understanding of the evolution process of de Sitter spacetime. In addition, in the evolution process of de Sitter spacetime, the role of each parameter on the evolution of spacetime is a matter of great concern. In recent years, with the study of dark energy, the study of the thermodynamic properties of de Sitter spacetime has attracted people's attention. The study of Hawking radiation in de Sitter spacetime shows that when the parameters of de Sitter spacetime satisfy certain conditions, spacetime has a black hole horizon and a cosmological horizon, both horizons have Hawking radiation, and the radiation temperatures of two horizons are generally different, so spacetime does not meet the requirements of thermal equilibrium stability, which brings certain difficulties to the study of the thermodynamic properties of de Sitter spacetime.
The aim of this paper is to study the thermodynamic behavior of asymptotically dS black holes and, in particular, to investigate if such black holes admit phase transitions similar to AdS cousins \cite{David Kubiznak12,Rong-Gen Cai13,Sharmila Gunasekaran12,Antonia M. Frassino14,Natacha Altamirano14,Natacha Altamirano13,Joy Das Bairagya20,X. H. Ge15,Rong-Gen Cai16,Jia-Lin Zhang15b,Ren Zhao13,Ren Zhao13b,Shao-Wen Wei09,S H Hendi17,A. Dehghani20,Rabin Banerjee20,M. Chabab2019,M. Chabab2016,Meng-Sen Ma17b,Dehyadegari2020,Daniela Mago2020,Sajadi2019,Hendi2019,Hendi2018,Eslam,Hendi2016a,Hendi2016b,Zou17,Jianfei15}, one approach to such a task is that of the effective temperature. An observer  is located in an observable part of the universe, between the black hole horizon and the cosmological horizon. For such an observer a new effective first law is imposed such that the system is assigned a total entropy that equals the sum of the black hole horizon and the cosmological horizon entropies, see e.g.~\cite{Yuichi Sekiwa06,Urano09}

\begin{equation}\label{1.3}
S=S_{+}+S_{c},
\end{equation}
where $S_{+}$ and $S_{c}$ are the entropy of the black hole horizon and the cosmological horizon, respectively, according to this, people gave the effective thermodynamic quantities of different de Sitter spacetime, and studied the critical behavior of spacetime. They obtained that the de Sitter spacetime also has a phase transition similar to AdS black holes~\cite{Yuichi Sekiwa06,Urano09,Saoussen19,David Kubiznak16,Sumarna Haroon19,Fil Simovic08,Chabab,Brian13,Bhattacharya16,McInerney,Panagiota17}. However, the de Sitter spacetime thermodynamic system is composed of the thermodynamic subsystems corresponding to the black hole horizon and the cosmological horizon, respectively. The radiation temperature of two subsystems is generally different. The total entropy of the thermodynamic system composed of two subsystems with different temperatures is sum of the entropy of the subsystems, this expression is imperfect. Firstly, it is reflected in the effective temperature derived from it \cite{Yuichi Sekiwa06,Urano09,Brian13}
\begin{equation}\label{1.4}
\it{T_{eff}}=\frac{T_{+}T_{c}}{T_{+}+T_{c}},
\end{equation}
for the effective temperature given by Eq.~(\ref{1.4}), when the spacetime satisfies certain conditions \cite{Romans92}, the black hole horizon radiation temperature $\it{T}_{+}$ is equal to the cosmological horizon radiation temperature $\it{T_{c}}$, and the effective temperature of spacetime is $\it{T_{eff}}=\it{T}_{+}/2=\it{T_{c}}/2$ , this result is unacceptable. In recent literatures \cite{Li-Chun Zhang16,Li-Chun Zhang19,Yubo Ma2020,Xiong-Ying72}, people consider the interaction between the two horizons, the entropy of RN-dS spacetime is sum of the corresponding entropy of two horizons plus the interaction term, namely
\begin{equation}\label{1.5}
\it{S}=\it{S_{c}}(1+x^{2}+f_{0}(x)),
\end{equation}
where $x$ is the ratio of the two horizons
\begin{equation}\label{1.6}
f_{0}(x)=\frac{8}{5}(1-x^{3})^{2/3}-\frac{2(4-5x^{3}-x^{5})}{5(1-x^{3})},
\end{equation}
thus, we obtain the effective temperature, when the temperatures of the two horizons are equal, the effective temperature is $\it{T_{eff}}=\it{T}_{+}=\it{T_{c}}$ , this result meets the requirements of the ordinary thermodynamic system, therefore, when considering the interaction between the two horizons, the space-time effective thermodynamic quantity obtained has more universal physical meaning.

Although people have obtained gratifying results in the thermal properties of various AdS black holes and the dS spacetime, theoretically, taking AdS space-time black holes and dS spacetime as a thermodynamic system, naturally there are critical behaviors and phase transition processes, but so far, the statistical mechanics background of black holes as a thermodynamic system is not clear. Therefore, it is very meaningful to study the relationship between the thermodynamic properties of various AdS spacetime and dS spacetime black holes. Exploring the internal relationship between the thermodynamics of different AdS spacetime and dS spacetime black holes not only helps to further understand the entropy, temperature, heat capacity and other properties of black holes, but also has important significance for perfecting the self-consistent black hole thermodynamic geometry theory.
In this paper, on the basis of considering the interaction between the black hole horizon and the cosmological horizon to obtain the effective thermodynamic quantity of spacetime~\cite{Li-Chun Zhang16,Li-Chun Zhang19,Yubo Ma2020,Xiong-Ying72}, the thermodynamic properties of RN-dS spacetime are discussed. The spacetime of RN-dS has the continuous phase transition which is similar to Van de Waals system or charged AdS black hole. The position where the continuous phase transition occurs in the RN-dS spacetime is related to the electric potential of the black hole horizon. This phenomenon reflects that in the process of the RN-dS spacetime evolves to a pure dS spacetime, due to the different electric potential at the horizon of the black hole, the evolution process of spacetime is different. Therefore, the electric potential at the horizon of the black hole plays a key role in the evolution of spacetime. The results of theoretical research on the evolution of spacetime of RN-dS can be used to simulate the evolution of universe, opening a new way to find the internal reasons that promote the accelerated expansion of universe.

This article is structured as follows, for the convenience of discussion, in section 2, the effective thermodynamic quantity of RN-dS spacetime is reviewed, and the change of the space-time effective thermodynamic quantity with the position of the two horizons is discussed. In section 3, the method of studying the phase transition of the Van de Waals system or the charged AdS black hole is extended to the study of RN-dS spacetime, and the RN-dS spacetime has a phase transition similar to the Van de waals system or the charged AdS black hole. According to Ehrenfest's classification method of phase transitions, it is obtained that RN-dS spacetime continuous first-order phase transition, and the phase transition point is determined by the electric potential at the black hole horizon. In section 4, in order to explore the microscopic state inside the black hole, the Landau continuous phase transition theory is used to discuss the critical phenomena and critical exponents of RN-dS spacetime and find that the order parameters are similar to those in ferromagnetic systems. In section 5, we analyze the nature of this phase transition using Ehrenfest's scheme, and the conclusion is given In section 6. (we use the units $G=\hbar=k_B=c=1$ )

\section{Effective thermodynamic quantity of RN-dS spacetime}
The static spherically symmetric solution of Einstein equation for a RN-dS black hole matters have been obtained

\begin{equation}\label{2.1}
d s^{2}=-g(r)d t^{2}+g^{-1}d r^{2}+r^{2}d\Omega^{2}_{2},
\end{equation}
with the horizon function
\begin{equation}\label{2.2}
g(r)=1-\frac{2M}{r}+\frac{Q^{2}}{r^{2}}-\frac{r^{2}}{l^{2}},
\end{equation}
whre $M$ and $Q$ are the black hole mass and charge, $l$ is the curvature radius of dS pace. Black hole horizon position $r_{+}$ and the position of the cosmological horizon $r_{c}$ satisfy $g(r_{+,c})=0$.
Considering the connection between the black hole horizon and the cosmological horizon, we can derive the effective thermodynamic quantities and corresponding first law of black hole thermodynamics [55-58]
\begin{equation}\label{2.3}
d M=\it{T_{eff}} d S-\it{P_{eff}} d V+\it{\phi_{eff}} d Q,
\end{equation}
here the thermodynamic volume is that between the black hole horizon and the cosmological horizon, namely~\cite{Sharmila Gunasekaran12,Panagiota17,Li-Chun Zhang16}
\begin{equation}\label{2.4}
V=\frac{4\pi}{3}(r_{c}^{3}-r_{+}^{3})=\frac{4\pi}{3x^{3}}r_{+}^{3}(1-x^{3}),
\end{equation}
where, $x=r_{+}/r_{c}$ is the ratio of the black hole horizon position $r_+$ and the position of the cosmological horizon $r_c$ .
Considering that the entropy corresponding to the two horizons is only an explicit function
of the position of the horizon, we set the total entropy of the system as

\begin{equation}\label{2.5}
S=\pi r_{c}^{2}(1+x^{2}+f_{0}(x))=\pi r_{+}^{2}(1+x^{2}+f_{0}(x))/\it{x^{2}},
\end{equation}
here the undefined function $f(x)$ represents the extra contribution from the correlations of the two horizons. Substitute Eq.~(\ref{2.4}) and (\ref{2.5}) into Eq.~(\ref{2.3}), we can obtain the effective temperature $T_{e\!f\!f}$ of the system, effective pressure $\it{P_{eff}}$ and $f_{0}(x)$ are as follows~\cite{Li-Chun Zhang16,Li-Chun Zhang19}
\begin{equation}\label{2.6}
\it{T_{eff}}=\frac{1-x}{4 \pi r_{+}(1+x^{4})}\{[(1+x)(1+x^{3})-2x^{2}]-\frac{Q^{2}}{r_{+}^{2}}[(1+x+x^{2})(1+x^{4})-2x^{3}]\},
\end{equation}
\begin{equation}\label{2.7}
\begin{split}
\it{P_{eff}}=&\frac{x(1-x)}{8 \pi r_{+}^{2}(1+x^{4})}[(\frac{1}{1+x+x^{2}})\left((1+2x)-\frac{Q^{2}(1+2x+3x^{2})}{r_{+}^{2}}\right)(1+x^{2}+f_{0}(x))\\
&-(1+x)x(1-\frac{Q^{2}(1+x^{2})}{r_{+}^{2}})(x+f'_{0}(x)/2)],
\end{split}
\end{equation}
where
\begin{equation}\label{2.8}
f_{0}(x)=-\frac{2\left(4-5 x^{3}-x^{5}\right)}{5\left(1-x^{3}\right)}+\frac{8}{5}\left(1-x^{3}\right)^{2 / 3}.
\end{equation}

When spacetime effective temperature $\it{T_{eff}}$ is given, we can obtain the inverse of the black hole horizon position $y=\frac{1}{r_{+}}$ from Eq.~(\ref{2.6}) as follows

\begin{equation}\label{2.9}
y_{1}=2\sqrt{\frac{f_{2}(x)}{3Q^{2}f_{3}(x)}}\cos\theta, \quad y_{2}=2\sqrt{\frac{f_{2}(x)}{3Q^{2}f_{3}(x)}}\cos(\theta+120^{\circ}),\quad y_{3}=2\sqrt{\frac{f_{2}(x)}{3Q^{2}f_{3}(x)}}\cos(\theta+240^{\circ}),
\end{equation}
where
\begin{equation}\label{2.10}
\begin{split}
f_{1}(x)=&\frac{4\pi(1+x^{4})}{1-x}, \quad f_{2}(x)=(1+x)(1+x^{3})-2x^{2},\quad f_{3}(x)=(1+x+x^{2})(1+x^{4})-2x^{3},\\
&\theta=\frac{1}{3}\arccos\left( -\frac{\it{T_{eff}}f_{1}(x)}{2}\sqrt{\frac{27Q^{2}f_{3}(x)}{f_{2}^{3}(x)}}\right).
\end{split}
\end{equation}
Since $y_{2}$ is a negative value, there is no physical meaning.

\begin{figure}[htp]
\centering
\subfigure[]{\includegraphics[width=0.45\textwidth]{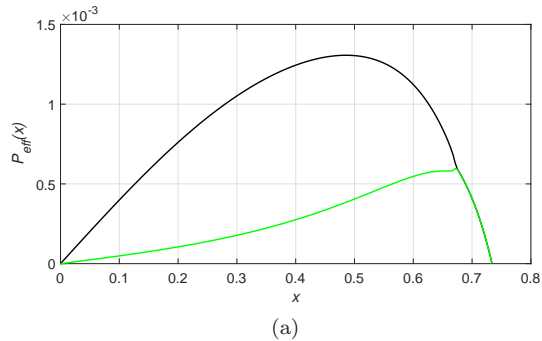}}
\caption{$\it{P_{eff}}$ varies with the ratio $x$ of two horizons.}
\label{fig2.1}
\end{figure}

Given the effective temperature $\it{T_{eff}}$ , we can obtain two different black hole horizon positions from Eq.~(\ref{2.9}). When the different black hole horizon positions of Eq.~(\ref{2.9}) are taken, they correspond to different $\it{P_{eff}}(x)-x$ curves under isothermal conditions as shown in Fig.~\ref{fig2.1}, i.e., when the effective temperature $\it{T_{eff}}(x)$ of spacetime and the position ratio of the two horizons are equal, the spacetime corresponds to two different pressures $\it{P_{eff}}(x)$.

According to the requirements of the stable equilibrium of the thermodynamic system, when the system is to reach the thermodynamic equilibrium state, the material with the higher internal pressure will flow to the lower part, so that the system can reach equilibrium. Therefore, the spacetime can only be in the phase with lower pressure when the effective temperature $\it{T_{eff}}(x)$ is the same with the position $x$ of the two horizons, that is, when the effective temperature $\it{T_{eff}}(x)$ is set, the effective pressure $\it{P_{eff}}(x)$ of spacetime evolves along the curve of lower pressure. Therefore, when we discuss the thermodynamic properties of spacetime, we only consider $y_{3}$ in Eq.~(\ref{2.9}) .

When the effective pressure $\it{P_{eff}}$ of the spacetime is given, we can obtain the horizon position $r_{+}^{2}=R$ of the black hole from Eq.~(\ref{2.7}), satisfying
\begin{equation}\label{2.11}
R=\frac{-F_{2}(x)+\sqrt{F^{2}_{2}(x)-4F_{1}(x)F_{3}(x)}}{2F_{1}(x)},
\end{equation}
where
\begin{align}\label{2.12}
F_{1}(x)&=\it{P_{eff}}\frac{8\pi(1+x^{4})}{x(1-x)},\quad F_{2}(x)=(1+x)x(x+f'(x)/2)-\frac{(1+2x)(1+x^{2}+f(x))}{1+x+x^{2}},\nonumber\\
F_{3}(x)&=Q^{2}\left(\frac{(1+2x+3x^{2})(1+x^{2}+f(x))}{1+x+x^{2}}-(1+x)x(1+x^{2})(x+f'(x)/2)\right).
\end{align}

\section{RN-dS spacetime phase transition}
In this section, we would like to study the phase transition in the canonical ensemble. From Eq.~(\ref{2.4})-(\ref{2.7}), we can obtain the equation of state reflecting the thermodynamic characteristics of RN-dS spacetime
\begin{equation}\label{3.1}
f(\it{T_{eff}}, \it{P_{eff}}, V, Q)=0.
\end{equation}
The Gibbs free energy of RN-dS spacetime  is
\begin{equation}\label{3.2}
\begin{aligned}
G(r_{+},x)&=M-\it{T_{eff}}S+\it{P_{eff}}V=\frac{r_{+}(1+x)}{2(1+x+x^{2})}+\frac{Q^{2}(1+x)(1+x^{2})}{2r_{+}(1+x+x^{2})}\\
&-r_{+}\frac{[(1+x^{2}+f(x))/x^{2}](1-x)}{4(1+x^{4})}\{(1+x)(1+x^{3})-2x^{2}-\frac{Q^{2}}{r_{+}^{2}}[(1+x+x^{2})(1+x^{4})-2x^{3}]\}\\
&-r_{+}\frac{x(1-x)(1-x^{3})}{6x^{2}(1+x^{4})}[\frac{1}{1+x+x^{2}}((1+2x)-\frac{Q^{2}(1+2x+3x^{2})}{r_{+}^{2}})(1+x^{2}+f_{0}(x))\\
&-(1+x)x(1-\frac{Q^{2}(1+x^{2})}{r_{+}^{2}})(x+f_{0}'(x)/2)].
\end{aligned}
\end{equation}
The critical point denoting a second-order phase transition is determined by the following equation
\begin{equation}\label{3.3}
\left(\frac{\partial \it{P_{eff}}}{\partial V}\right)_{\it{T_{eff}},Q}=\left(\frac{\partial^{2} \it{P_{eff}}}{\partial V^{2}}\right)_{\it{T_{eff}},Q}=0.
\end{equation}
From Eq.~(\ref{3.3}), we can obtain that the critical point position of spacetime satisfies $r_{+}^{c}=2.64432Q, x^{c}=0.656434$, When $Q=1, \it{T_{eff}^{c}}=\frac{0.2869}{4\pi r_{+}^{c}}=0.00863395, \it{P_{eff}^{c}}=\frac{0.10257633}{8\pi (r_{+}^{c})^{2}}=0.000583686, V_{c}=\frac{4\pi}{3 x^{3}}(r_{+}^{c})^{3}(1-x^{3})=196.363483$.

\begin{figure}[htp]
\centering
\subfigure[]{\includegraphics[width=0.45\textwidth]{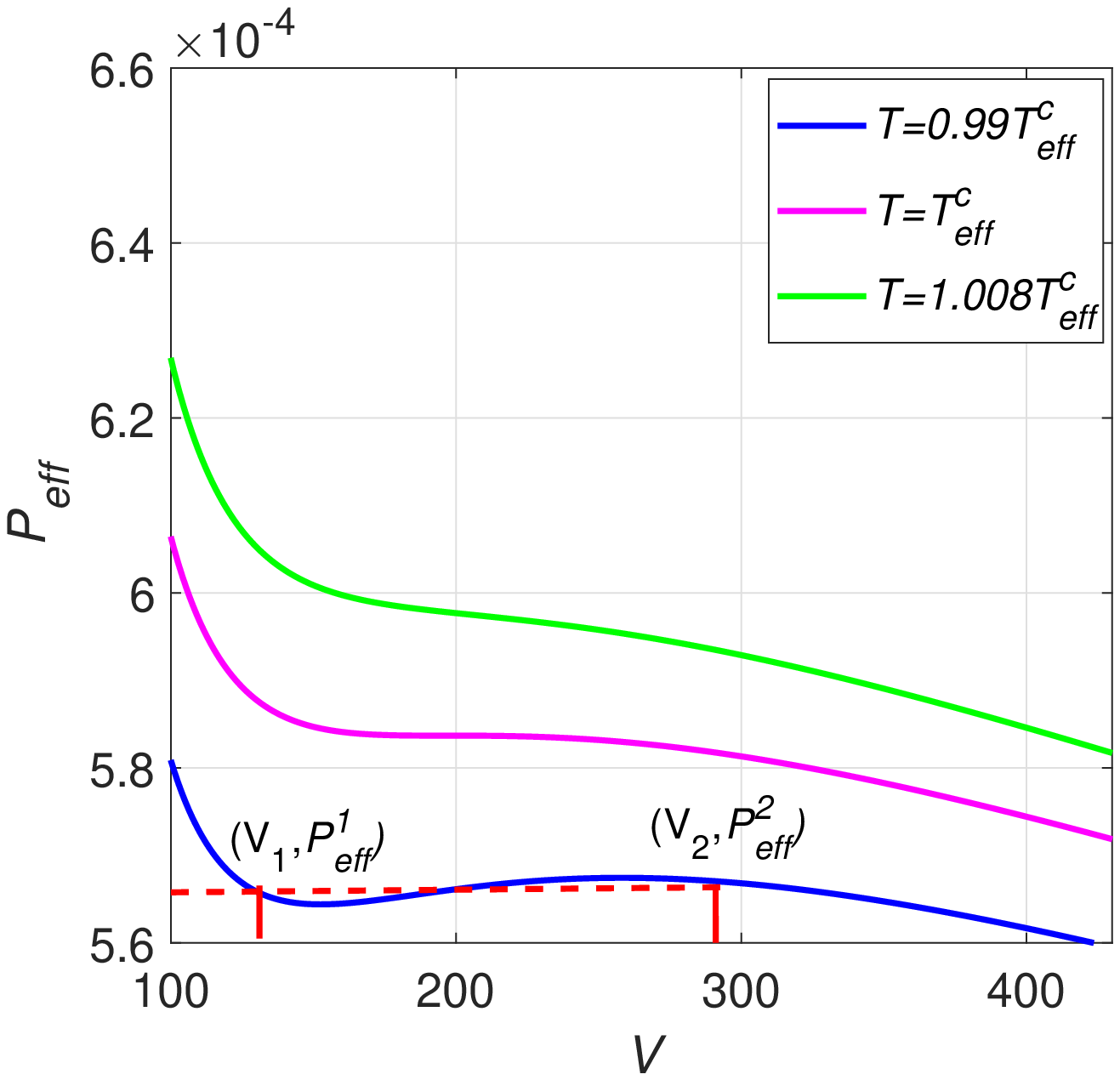}}
\caption{The isotherm $\it{P_{eff}}(x)-V(x)$ curves.}
\label{fig3.1}
\end{figure}

From Eq.~(\ref{2.4}), (\ref{2.6}) and (\ref{2.7}), we can draw the isotherm $\it{P_{eff}}(x)-V(x)$ curves as shown in Fig.~\ref{fig3.1}

The isotherms of the state equation that satisfying Eq.~(\ref{3.1}) is in the range $\it{P_{eff}}^{1}<\it{P_{eff}}<\it{P_{eff}}^{2}$. There are three possible $V$ values corresponding to a $\it{P_{eff}}$ value, as shown in Fig.~\ref{fig3.1}. In the range of $V_{1}<V<V_{2}$, we have $\left(\frac{\partial \it{P_{eff}}}{\partial V}\right)_{\it{T_{eff},Q}}>0$ ,this does not meet the requirements of the equilibrium stability of the thermodynamic system, these states cannot be realized as a uniform system. According to the minimum requirement of the Gibbs function, in the range of $\it{P_{eff}}^{1}<\it{P_{eff}}<\it{P_{eff}}^{2}$, a dashed line is used to replace the isotherm curve satisfying Eq.~(\ref{3.1}). The area of the two parts enclosed by the red dashed line and the curve is equal, that is, Maxwell equal area law. With the effective temperature increases, the area of the two parts gradually decreases. When the area approaches zero, the three intersections of the straight line and the curve tend to one point, which is the critical point. The position of the critical point satisfies the Eq.~(\ref{3.3}).

\begin{figure}[htp]
\centering
\subfigure[]{\includegraphics[width=0.45\textwidth]{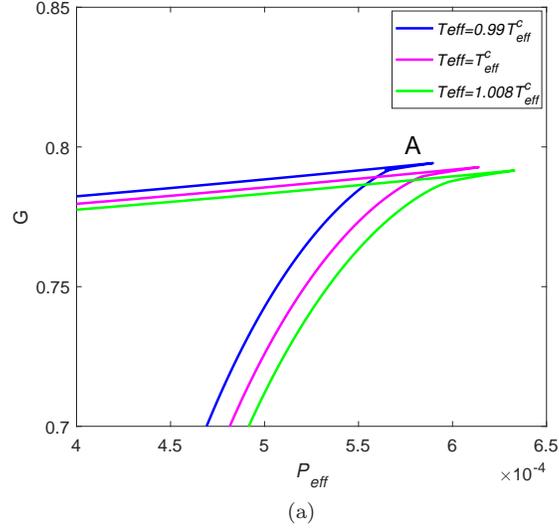}}
\caption{The $G-\it{P_{eff}}$ curves under the effective temperature.}
\label{fig3.2}
\end{figure}

From Eq.~(\ref{2.7}) and Eq.~(\ref{3.2}), we can draw the isotherm $G-\it{P_{eff}}$ curves as shown in Fig.~\ref{fig3.2},
we can see that when the effective temperature satisfies $\it{T_{eff}}>\it{T_{eff}}^{c}$, the single value of the isotherm curve changes continuously, and when the effective temperature is $\it{T_{eff}}<\it{T_{eff}}^{c}$, the $G-\it{P_{eff}}$ curve has an intersection point A, this is similar to the Van de Waals system and the AdS black hole. According to the Gibbs free energy criterion under the isothermal and pressure conditions of the system, the first-order phase transition occurs at point A of the $G-\it{P_{eff}}$ curve in spacetime. (Point A is marked at the intersection of the two curves on the temperature curve $\it{T_{eff}}<\it{T_{eff}}^{c}$)
The heat capacity at constant pressure is as follows

\begin{equation}\label{3.4}
\begin{aligned}
C_{\it{P_{eff}},Q} &=\it{T_{eff}}\left(\frac{\partial S}{\partial \it{T_{eff}}}\right)_{\it{P_{eff}},Q}=\it{T_{eff}} \frac{\frac{\partial S}{\partial x}\frac{\partial \it{P_{eff}}}{\partial r_{+}}-\frac{\partial S}{\partial r_{+}}\frac{\partial \it{P_{eff}}}{\partial x}}{\frac{\partial \it{T_{eff}}}{\partial x}\frac{\partial \it{P_{eff}}}{\partial r_{+}}-\frac{\partial \it{P_{eff}}}{\partial r_{+}}\frac{\partial \it{T_{eff}}}{\partial x}}.
\end{aligned}
\end{equation}

\begin{figure}[htp]
\centering
\subfigure[]{\includegraphics[width=0.45\textwidth]{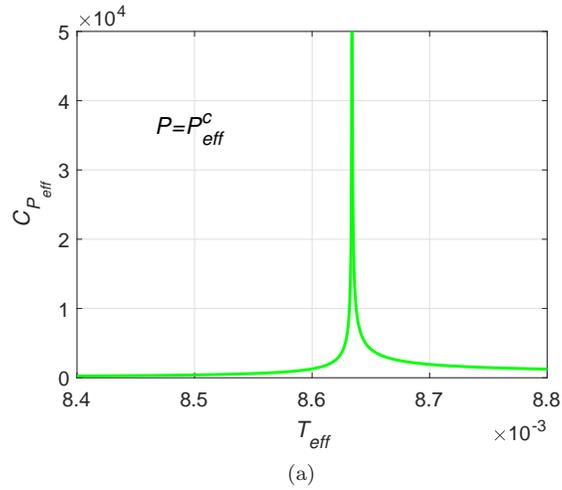}}
\caption{The isobaric $C_{\it{P_{eff}}}-\it{T_{eff}}$ curve.}
\label{fig3.3}
\end{figure}

From Eqs.~(\ref{2.6}),~(\ref{2.7}) and~(\ref{3.4}), we can draw the  $C_{\it{P_{eff}}}-\it{T_{eff}}$ curve as shown in Fig.~\ref{fig3.3}.

Volume expansion coefficient is as follows
\begin{equation}\label{3.5}
\begin{aligned}
\beta&=\frac{1}{V}\left(\frac{\partial V}{\partial \it{T_{eff}}}\right)_{\it{\it{P_{eff},Q}}}=\frac{1}{V} \frac{\frac{\partial V}{\partial x}\frac{\partial \it{P_{eff}}}{\partial r_{+}}-\frac{\partial V}{\partial r_{+}}\frac{\partial \it{P_{eff}}}{\partial x}}{\frac{\partial \it{T_{eff}}}{\partial x}\frac{\partial \it{P_{eff}}}{\partial r_{+}}-\frac{\partial \it{T_{eff}}}{\partial r_{+}}\frac{\partial \it{P_{eff}}}{\partial x}}.
\end{aligned}
\end{equation}

\begin{figure}[htp]
\centering
\subfigure[]{\includegraphics[width=0.45\textwidth]{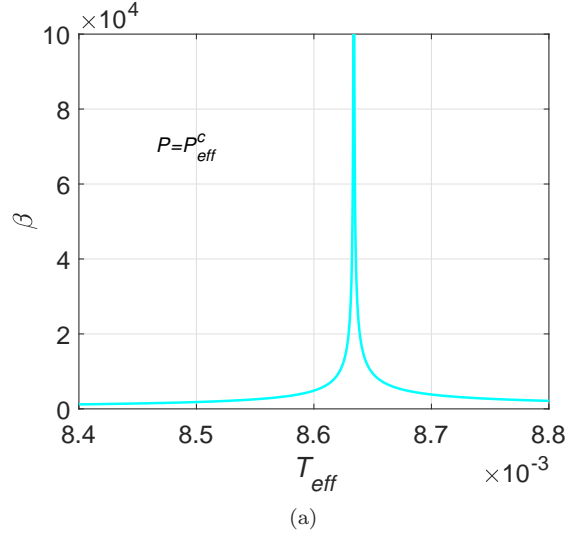}} %
\caption{The isobaric $\beta-\it{T_{eff}}$ curve.}
\label{fig3.4}
\end{figure}

From Eqs.~(\ref{2.6}),~(\ref{2.7}) and~(\ref{3.5}), we can draw the heat capacity at constant pressure curve $\beta-\it{T_{eff}}$ as shown in Fig.~\ref{fig3.4}

Isothermal compression coefficient is

\begin{equation}\label{3.6}
\begin{aligned}
\kappa_{\it{T_{eff}}}=\frac{1}{V}\left(\frac{\partial V}{\partial \it{P_{eff}}}\right)_{\it{\it{T_{eff},Q}}}=\frac{1}{V} \frac{\frac{\partial V}{\partial x}\frac{\partial \it{T_{eff}}}{\partial r_{+}}-\frac{\partial V}{\partial r_{+}}\frac{\partial \it{T_{eff}}}{\partial x}}{\frac{\partial \it{T_{eff}}}{\partial x}\frac{\partial \it{P_{eff}}}{\partial r_{+}}-\frac{\partial \it{T_{eff}}}{\partial r_{+}}\frac{\partial \it{P_{eff}}}{\partial x}}.
\end{aligned}
\end{equation}

\begin{figure}[htp]
\centering
\subfigure[]{\includegraphics[width=0.45\textwidth]{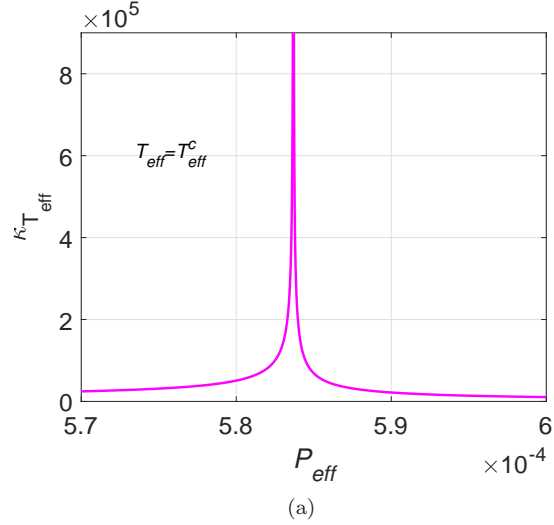}} %
\caption{The isotherm $\kappa_{\it{T_{eff}}}-\it{P_{eff}}$ curve.}
\label{fig3.5}
\end{figure}

From Eqs.~(\ref{2.6}),~(\ref{2.7}) and~(\ref{3.6}), we can draw the isothermal curve $\kappa_{\it{T_{eff}}}-\it{P_{eff}}$ as shown in Fig.~\ref{fig3.5}.

\begin{figure}[htp]
\centering
\subfigure[]{\includegraphics[width=0.45\textwidth]{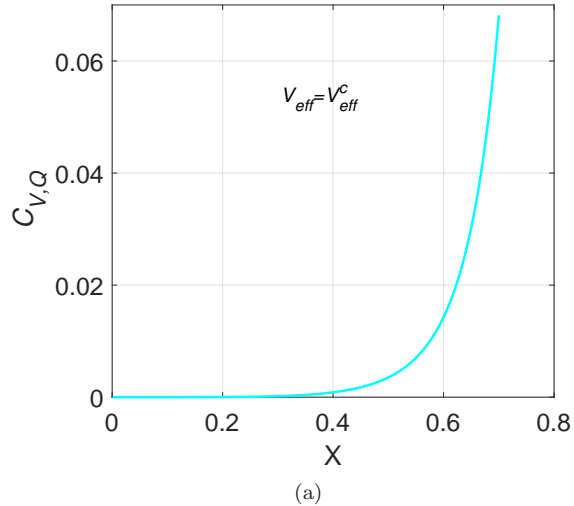}} %
\caption{The constant volume $C_{V,Q}-x$ curve.}
\label{fig3.6}
\end{figure}

The heat capacity at constant volume can be derived as
\begin{equation}\label{3.7}
\begin{aligned}
C_{V,Q}&=\it{T_{eff}}\left(\frac{\partial S}{\partial \it{T_{eff}}}\right)_{V,Q}=\it{T_{eff}} \frac{\frac{\partial S}{\partial x}\frac{\partial V}{\partial r_{+}}-\frac{\partial S}{\partial r_{+}}\frac{\partial V}{\partial x}}{\frac{\partial \it{T_{eff}}}{\partial x}\frac{\partial V}{\partial r_{+}}-\frac{\partial \it{T_{eff}}}{\partial r_{+}}\frac{\partial V}{\partial x}}.
\end{aligned}
\end{equation}
From Eqs.~(\ref{2.4}) and~(\ref{3.7}), we can draw $C_{V,Q}-x$ curve as shown in Fig.~\ref{fig3.6}.
It can be seen that the heat capacity at constant volume is not zero, while for the spherical symmetric AdS spacetime the heat capacity at constant volume is zero~\cite{Shao-Wen19,Shao-Wen2020,Run Zhou08,Yu-Xiao20}. Generally, for a universal thermodynamic system, the heat capacity at constant volume is not zero. Therefore, our results are more universal.
According to Ehrenfest's classification of phase transitions, if the Gibbs function and its first partial derivative in two-phase are continuous at the phase transition point, but there is a sudden change in the second partial derivative of the Gibbs function, it is called the second-order phase transition. From the above analysis, we know that when the effective temperature satisfies $\it{T_{eff}}<\it{T_{eff}^{c}}$, the entropy and thermodynamic volume of spacetime undergo a sudden change, and the first-order phase transition occurs in spacetime. When the effective temperature is $\it{T_{eff}}=\it{T_{eff}^{c}}$, the entropy and thermodynamic volume of spacetime are continuous, Fig.~\ref{fig3.3} -- Fig.~\ref{fig3.5} show that there are sudden changes in the heat capacity at constant pressure,  expansion coefficient and isothermal compression coefficient of space-time. The second-order phase transition occurs in space-time, and the mutation point is the second-order phase transition point.
It can be seen from Fig.~\ref{fig3.6} that for RN-dS spacetime, when we consider the interaction between the two horizons, the change curve of the  heat capacity at equal volume of the spacetime with the position ratio $x$ of the two horizons is not zero, that is, when the spacetime thermodynamic volume remains unchanged, the heat capacity is not zero. This is different from the  heat capacity at equal volume of spherically symmetric AdS spacetime which is zero\cite{Run Zhou08,Yu-Xiao20,Yan-Gang18,Yan-Gang2019}. However, in general thermodynamic systems, when the volume of the system remains unchanged, the heat capacity of the system is not zero. Thus, when considering the interaction of the two horizons, the obtained effective thermodynamic quantity and the spacetime effective thermodynamic quantity satisfy the equation of state closer to the characteristics of the ordinary thermodynamic system, so the thermodynamic system described by the effective thermodynamic quantity in spacetime is more universal.

\section{The critical phenomena and critical exponents of RN-dS spacetime}

Through the discussion of the space-time thermodynamic properties of RN-dS in the third part, due to the continuity of Gibbs free energy, we found that the space-time has characteristics and the second-order phase transition similar to that of the Van de Waals gas-liquid two-phase transition and the AdS black hole~\cite{Hendi2018,Eslam,Hendi2016a,Hendi2016b,Xiong-Ying20,Xiong-Ying2019}.
In the field of the critical point of continuous phase transition, the  heat capacity at constant pressure, the expansion coefficient and the isothermal compression coefficient, etc., have a jump or infinite spike. People use power functions to express the characteristics of some critical areas, and its power is called the critical exponent. People have found that various ferromagnetic systems have the following common experimental laws in the field of critical points: when $t=T-T_{c}$, the changes of spontaneous magnetization satisfying the following law

\begin{equation}\label{4.1}
M\propto (-t)^{\beta}, \quad t\rightarrow-0.
\end{equation}
When above the critical temperature, spontaneous magnetization is zero (i.e., $M=0$). The magnetic susceptibility $\chi$ of various ferromagnetic materials varies with $t\rightarrow \pm0$ satisfying the following law

\begin{equation}\label{4.2}
\chi \propto t^{-\gamma}, \quad t\rightarrow+0; \quad \chi \propto (-t)^{-\gamma'}, \quad t\rightarrow-0.
\end{equation}
when $t\rightarrow 0$ , the relationship between magnetization and external magnetic field is
\begin{equation}\label{4.3}
M\propto H^{1/\delta}.
\end{equation}
The zero-field heat capacity of ferromagnetic materials obeys the following law
\begin{equation}\label{4.4}
C_{H} \propto t^{-\alpha}, \quad t>0; \quad C_{H} \propto (-t)^{-\alpha'}, \quad t<0.
\end{equation}

 It is useful to introduce the concept of critical exponent. When approaching the critical point, various thermodynamic functions either diverge, or tend to zero, or remain limited. It is obtained that not only the fluid systems but various ferromagnetic systems follow the same law near the critical point, and when the liquid-gas density difference $\rho_{l}-\rho_{g}$ is compared to the magnetization $M$, the pressure $P$ is compared to the magnetic field strength $H$, and the isothermal compression coefficient $\kappa_{\tau}$ is compared to the magnetic susceptibility $\chi$, although the physical characteristics of the two systems are very different, the behavior in the critical point is extremely similar. Not only is the changing law same, but the critical exponent is also roughly the same. This fact shows that the critical phenomenon has a certain universality.

The critical exponent of the AdS black hole system is consistent with the theoretical calculation results of the critical exponent describing the critical behavior of the ferromagnetic system \cite{Shao-Wen2020} as follows

\begin{equation}\label{4.5}
\beta=1/2, \quad \alpha=\alpha'=0, \quad \gamma=\gamma'=1, \quad \delta=3.
\end{equation}

The phenomenon of RN-dS spacetime in the critical point domain is our concern. The thermodynamic properties of dS spacetime are not the same as the Van de Waals system, nor the same as the ferromagnetic system. Its phase transition depends on the position of the two horizons. Therefore, the discussion about the critical phenomenon of dS spacetime will help us to further understand the thermodynamic properties of dS spacetime. According to Landau theory of phase transition, the phase transition occurs due to the change of the degree of order within the system. In order to describe the ordering in the system, we can introduce the order parameter. For dS spacetime, we introduce the order parameter $\eta$ as a generalized coordinate, then Gibbs free energy is the function of $T, P$ and $\eta$. We mainly consider the relationship between $g$ and $\eta$, in the vicinity of $\eta\approx 0$, the power series of $G$ expansion, to the fourth power is as follows (for convenience, we will write $\it{T_{eff}}$ as $T$, and $\it{P_{eff}}$ as $P$ )

\begin{equation}\label{4.6}
G=G_{0}(T, P)+a \eta+A \eta^{2}+b \eta^{3}+B \eta^{4},
\end{equation}
where $a, \eta, A, b$ and $B$ are the function of $T$ and $P$, it is nothing to do with $\eta$, and use $\xi$ to express the generalized force of $\eta$, thus
\begin{equation}\label{4.7}
\xi=\frac{\partial G}{\partial \eta}=a+2A \eta+3b \eta^{2}+4B \eta^{3},
\end{equation}
under the requirement of $\eta=0$ (i.e. disorder condition), there will be $a=0$, and when the symbol of $\eta$ is changed, the symbol of $\xi$ changes again. Under these conditions, Eq.~(\ref{4.6}) is simplified to

\begin{equation}\label{4.8}
G=G_{0}(T, P)+A(T, P) \eta^{2}+B(T, P) \eta^{4},
\end{equation}
and
\begin{equation}\label{4.9}
\xi=2 A \eta+4B \eta^{3},
\end{equation}
without external force, i.e., $\xi=0$, we have
\begin{equation}\label{4.10}
 A \eta+2B \eta^{3}=0,
\end{equation}
its three roots are
\begin{equation}\label{4.11}
\eta=0, \quad \eta=\pm\sqrt{-\frac{A}{2B}},
\end{equation}
if $B<0$, thus $\eta\rightarrow \infty$, it will lead to $G\rightarrow -\infty$, $U\rightarrow -\infty$, the system is unstable. A stable system must have $B>0$. For the case of $A>0$ in Eq.~(\ref{4.10}), it has only one real root $\eta=0$, according to Eq.~(\ref{4.7})
\begin{align}\label{4.12}
\frac{\partial G}{\partial \eta}=2A \eta+4B \eta^{3},\\
\frac{\partial^{2} G}{\partial \eta^{2}}=2A +12B \eta^{2},
\end{align}
the real root is $\eta=0$ under the condition of $A>0$, then $\frac{\partial G}{\partial \eta}=0$, $\frac{\partial^{2} G}{\partial \eta^{2}}>0$, so $G$ takes the minimum value, corresponding to a stable equilibrium. Under the condition $A<0$ with in Eq.~(\ref{4.10}), there are three real roots, for $\eta=0$, we have $\frac{\partial^{2} G}{\partial \eta^{2}}<0$, so $G$ takes the maximum value, corresponding to an unstable state. The other two roots $\eta=\pm\sqrt{-\frac{A}{2B}}$ make $\frac{\partial G}{\partial \eta}=0$, $\frac{\partial^{2} G}{\partial \eta^{2}}=8| {A}|>0$, thus $G$ is a minimum value, which corresponds to a stable state. It can be seen that the system will be in a disordered phase with high symmetry under the condition of $A>0$, and the system will be in an ordered phase with low symmetry under the condition of $A<0$. Phase transition occurs between each other for $A=0$, given by
\begin{equation}\label{4.13}
A(T_{c}, P_{c})=0,
\end{equation}
where $T_{c}$ and $P_{c}$ are the critical temperature and critical pressure respectively. $A(T, P)$ can be expanded with the power series near the critical point
\begin{equation}\label{4.14}
A(T, P)=\tilde{a}(T_{c}, P_{c})(T-T_{c})+\tilde{b}(T_{c}, P_{c})(P-P_{c}),
\end{equation}
where $\tilde{a}(T_{c}, P_{c})=(\frac{\partial A}{\partial T})_{T_{c}, P_{c}}$, $\tilde{b}(T_{c}, P_{c})=(\frac{\partial A}{\partial P})_{T_{c}, P_{c}}$. Taking $\tilde{a}>0$, $\tilde{b}>0$, means that the phase of $T>T_{c}$, $P=P_{c}$ is a stable state, that is the phase of disorder, and the phase of $T<T_{c}$, $P=P_{c}$ is an ordered phase. Substituting Eq.~(\ref{4.14}) into Eq.~(\ref{4.8}) gives
\begin{equation}\label{4.15}
G=G_{0}(T, P)+[\tilde{a}(T_{c}, P_{c})(T-T_{c})+\tilde{b}(T_{c}, P_{c})(P-P_{c})]\eta^{2}.
\end{equation}
From Eq.~(\ref{4.15}), we can see that Gibbs free energy $G$ is continuous at the critical point. Entropy of the system is
\begin{equation}\label{4.16}
S=-\frac{\partial G}{\partial T}=S_{0}-\tilde{a}(T_{c}, P_{c})\eta^{2},
\end{equation}
and the volume
\begin{equation}\label{4.17}
V=\frac{\partial G}{\partial P}=V_{0}+\tilde{b}(T_{c}, P_{c})\eta^{2},
\end{equation}
is continuous. When $T\geq T_{c}$, $\eta=0$, $S=S_{0}$, when $T\leq T_{c}$ ,
\begin{equation}\label{4.18}
\eta=\pm\sqrt{\frac{\tilde{a}(T_{c}, P_{c})(T_{c}-T)+\tilde{b}(T_{c}, P_{c})(P_{c}-P)}{2B}},
\end{equation}
thus
\begin{align}\label{4.19}
S&=S_{0}+\tilde{a}(T_{c}, P_{c})\frac{\tilde{a}(T_{c}, P_{c})(T-T_{c})+\tilde{b}(T_{c}, P_{c})(P-P_{c})}{2B},
\end{align}
\begin{equation}\label{4.119}
V=\frac{\partial G}{\partial P}=V_{0}+\tilde{b}(T_{c}, P_{c})\frac{\tilde{a}(T_{c}, P_{c})(T-T_{c})+\tilde{b}(T_{c}, P_{c})(P-P_{c})}{2B}.
\end{equation}
From Eq.~(\ref{4.19}) and~(\ref{4.119}), the entropy $S$ and volume $V$ of the system are continuous at the critical point. Near the critical point the  heat capacity at constant pressure is
\begin{equation}\label{4.20}
C_{p}=T \left(\frac{\partial S}{\partial T}\right)_{p}=T_{c} \left(\frac{\partial S_{0}}{\partial T}\right)_{p}+\frac{\tilde{a}^{2}(T_{c}, P_{c})}{2B} T_{c}=C_{P0}+\frac{\tilde{a}^{2}(T_{c}, P_{c})}{2B} T_{c}.
\end{equation}
Expansion coefficient is
\begin{equation}\label{4.21}
\beta=\frac{1}{V} \left(\frac{\partial V}{\partial T}\right)_{p}=\beta_{0}+\frac{\tilde{a}(T_{c}, P_{c})\tilde{b}(T_{c}, P_{c})}{2V_{0}B}.
\end{equation}
Isothermal compression coefficient is
\begin{equation}\label{4.22}
\kappa_{T}=-\frac{1}{V} \left(\frac{\partial V}{\partial P}\right)_{T}=\kappa_{T}+\frac{\tilde{b}^{2}(T_{c}, P_{c})}{2V_{0}B}.
\end{equation}
The heat capacity at constant volume is
\begin{equation}\label{4.23}
C_{v}=T \left(\frac{\partial S}{\partial T}\right)_{T}=-T \left(\frac{\partial^{2} S}{\partial T^{2}}\right)=C_{V0}+\frac{\tilde{a}^{2}(T_{c}, P_{c})}{2B}{T_{c}},
\end{equation}
which has a sudden change at the point $T=T_{c}$, and the critical exponent is $\alpha=\alpha^{'}=0$.
From Eq.~(\ref{4.9})
\begin{equation}\label{4.24}
\chi=d \xi=2A d \eta+12B \eta^{2}d \eta,
\end{equation}
we obtain the generalized external force function whose order parameter is
\begin{equation}\label{4.25}
\chi=\frac{\partial \eta}{\partial \xi}=\frac{1}{2A +12B \eta^{2}}.
\end{equation}
When $P=P_{c}$, at high temperature $T>T_{c}, \eta=0$
\begin{equation}\label{4.26}
\chi=\frac{\partial \eta}{\partial \xi}=\frac{1}{2A}=\frac{1}{2\tilde{a}(T_{c}, P_{c})(T-T_{c})},
\end{equation}
when $T<T_{c}$
\begin{equation}\label{4.27}
\chi=\frac{\partial \eta}{\partial \xi}=-\frac{1}{4A}=\frac{1}{2\tilde{a}(T_{c}, P_{c})(T_{c}-T)},
\end{equation}
the critical exponent is $\gamma=\gamma'=1$. From Eq.~(\ref{4.18}), we have
\begin{align}\label{4.28}
\eta=0,  ~~~~~~~~~~~~~~~~~~~~~~~~~~~~~~~~ \quad T>T_{c},\\
 \eta=\pm \sqrt{\frac{\tilde{a}(T_{c}, P_{c})(T-T_{c})+\tilde{b}(T_{c}, P_{c})(P-P_{c})}{2B}}, \quad T<T_{c}.
\end{align}
the critical exponent is $\beta=1/2$. From Eq.~(\ref{4.9}) and Eq.~(\ref{4.14}), when $T=T_{c}, P=P_{c}, A(T, P)=0$
\begin{equation}\label{4.29}
\xi=4B\eta^{3},
\end{equation}
the critical exponent is $\delta=3$.

When Landau studied the second-order phase transition of the ferromagnetic system, the order parameter $\eta$ were taken as the spontaneous magnetization vector $M$. For RN-dS spacetime, the order parameter is related to the electric potential at the black hole horizon, because the critical point for the second-order phase transition in spacetime satisfies $r_{+}^{c}=2.64432Q$ and $x_{c}=0.656434$, the electric potential at the critical point of spacetime is a constant $Q/r_{+}^{c}=1/2.64432$. When the effective temperature of spacetime is $\it{T_{eff}}<\it{T_{eff}}^{c}$, it can be seen from the $\it{P_{eff}}-V$ isotherm curve that when the first-order phase transition occurs in spacetime, the black hole horizon corresponds to two different electric potentials $Q/r_{+1}$, $Q/r_{+2}$, the magnitude of $r_{+1}$ and $r_{+2}$ are determined by the effective temperature. Suppose $r_{+1}/r_{+2}=y$, when the first-order phase transition occurs in spacetime, the electric potential difference at the horizon of the black hole is
\begin{equation}\label{4.30}
\eta=\frac{Q}{r_{+2}}\frac{1-y}{y},
\end{equation}
where $r_{+2}$ is given by the Maxwell equal area law~\cite{Huaifan17,Huai-Fan111,Jun-Xin Zhao14,Hao Xu17,Belhaj15}. When the effective temperature approaches the critical temperature, we have $y\rightarrow 1, r_{+2}\rightarrow r_{+}^{c}$. When the time space effective temperature is $\it{T_{eff}}<\it{T_{eff}}^{c}$, we have $\eta=0$. When the effective temperature of the spacetime is $\it{T_{eff}}<\it{T_{eff}}^{c}$, satisfying the requirement of our order parameter.

\section{Phase transition From Ehrenfest's equations}
We now exploit Ehrenfest's scheme in order to understand the nature of the phase transition. Ehrenfest's scheme basically consists of a pair of equations known as Ehrenfest's equations of first and second kind. For a standard thermodynamic system these equations may be written as \cite{Banerjee004,Banerjee156,Niloofar Abbasvandi}
\begin{align}\label{5.1}
\left(\frac{\it{P_{eff}}}{\it{T_{eff}}}\right)_{s}&=\frac{C_{\it{P_{eff}}2}-C_{\it{P_{eff}}1}}{\it{T_{eff}}V(\beta_{2}-\beta_{1})}=\frac{\Delta C_{\it{P_{eff}}}}{\it{T_{eff}}V \Delta \beta},
\end{align}
\begin{align}\label{5.101}
\left(\frac{\it{P_{eff}}}{\it{T_{eff}}}\right)_{V}&=\frac{\beta_{2}-\beta_{1}}{\kappa_{\it{T_{eff}}2}-\kappa_{\it{T_{eff}}1}}=\frac{\Delta \beta}{\it{T_{eff}}V \Delta \kappa_{\it{T_{eff}}}}.
\end{align}
For a genuine second order phase transition both of these equations have to be satisfied simultaneously. From Eq.~(\ref{2.3}), we can obtain the following relation
\begin{equation}\label{5.2}
\left(\frac{\partial \it{P_{eff}}}{\partial \it{T_{eff}}}\right)_{s}=\left(\frac{\partial S}{\partial V}\right)_{\it{P_{eff}}}, \quad \left(\frac{\partial \it{P_{eff}}}{\partial \it{T_{eff}}}\right)_{V}=\left(\frac{\partial S}{\partial V}\right)_{\it{T_{eff}}},
\end{equation}
using Eq.~(\ref{5.2}) the Prigogine-Defay (PD) ratio $\Pi$ may be found to be
\begin{equation}\label{5.3}
\Pi=\left(\frac{\partial \it{P_{eff}}}{\partial \it{T_{eff}}}\right)_{S}/\left(\frac{\partial \it{P_{eff}}}{\partial \it{T_{eff}}}\right)_{V}=\left(\frac{\partial S}{\partial V}\right)_{\it{P_{eff}}}=\left(\frac{\partial S}{\partial V}\right)_{\it{T_{eff}}},
\end{equation}
at the critical point $(\it{T_{eff}}^{c}, \it{P_{eff}}^{c}, V^{c})$, satisfies the equation as follows
\begin{equation}\label{5.4}
\left(\frac{\partial \it{P_{eff}}}{\partial V}\right)_{\it{T_{eff}}}=0, \quad \left(\frac{\partial^{2} \it{P_{eff}}}{\partial V^{2}}\right)_{\it{T_{eff}}}=0,
\end{equation}
substituting Eq.~(\ref{5.2}) into Eq.~(\ref{5.1}) and~(\ref{5.101}), we can obtain
\begin{equation}\label{5.5}
\frac{\Delta C_{\it{P_{eff}}}}{\it{T_{eff}}^{c} V^{c} \Delta \alpha}=\left[\left(\frac{\partial S}{\partial V}\right)_{\it{P_{eff}}} \right]^{c}, \quad \frac{\Delta \alpha}{\Delta \kappa_{\it{T_{eff}}}}=\left(\frac{\partial S}{\partial V}\right)_{\it{P_{eff}}}^{c},
\end{equation}
apply Jacobi determinant for derivative transformation operation, then
\begin{align}\label{5.6}
(\frac{\partial S}{\partial V})_{\it{P_{eff}}}=\frac{\frac{\partial (S, \it{P_{eff}})}{\partial (x, r_{+})}}{\frac{\partial (V, \it{P_{eff}})}{\partial (x, r_{+})}}=\frac{(\frac{\partial S}{\partial x})_{r_{+}}(\frac{\partial \it{P_{eff}}}{\partial r_{+}})_{x}-(\frac{\partial S}{\partial r_{+}})_{x}(\frac{\partial P_{e\!f\!f}}{\partial x})_{r_{+}}}{(\frac{\partial V}{\partial x})_{r_{+}}(\frac{\partial \it{P_{eff}}}{\partial r_{+}})_{x}-(\frac{\partial V}{\partial r_{+}})_{x}(\frac{\partial \it{P_{eff}}}{\partial x})_{r_{+}}},\nonumber\\
(\frac{\partial S}{\partial V})_{\it{T_{eff}}}=\frac{\frac{\partial (S, \it{T_{eff}})}{\partial (x, r_{+})}}{\frac{\partial (V, \it{T_{eff}})}{\partial (x, r_{+})}}=\frac{(\frac{\partial S}{\partial x})_{r_{+}}(\frac{\partial \it{T_{eff}}}{\partial r_{+}})_{x}-(\frac{\partial S}{\partial r_{+}})_{x}(\frac{\partial \it{T_{eff}}}{\partial x})_{r_{+}}}{(\frac{\partial V}{\partial x})_{r_{+}}(\frac{\partial \it{T_{eff}}}{\partial r_{+}})_{x}-(\frac{\partial V}{\partial r_{+}})_{x}(\frac{\partial \it{T_{eff}}}{\partial x})_{r_{+}}}.
\end{align}
The equation that determines the position of the critical point can be expressed as follows
\begin{equation}\label{5.7}
\left(\frac{\partial \it{P_{eff}}}{\partial V}\right)_{\it{T_{eff}}}=\frac{\frac{\partial (\it{P_{eff}}, \it{T_{eff}})}{\partial (x, r_{+})}}{\frac{\partial (V, \it{T_{eff}})}{\partial (x, r_{+})}}=\frac{(\frac{\partial \it{P_{eff}}}{\partial x})_{r_{+}}(\frac{\partial \it{T_{eff}}}{\partial r_{+}})_{x}-(\frac{\partial \it{P_{eff}}}{\partial r_{+}})_{x}(\frac{\partial \it{T_{eff}}}{\partial x})_{r_{+}}}{(\frac{\partial V}{\partial x})_{r_{+}}(\frac{\partial \it{T_{eff}}}{\partial r_{+}})_{x}-(\frac{\partial V}{\partial r_{+}})_{x}(\frac{\partial \it{T_{eff}}}{\partial x})_{r_{+}}}=0,
\end{equation}
when $(\frac{\partial V}{\partial x})_{r_{+}}(\frac{\partial \it{T_{eff}}}{\partial r_{+}})_{x}-(\frac{\partial V}{\partial r_{+}})_{x}(\frac{\partial \it{T_{eff}}}{\partial x})_{r_{+}}\neq 0$, the critical point satisfies the following equation
\begin{equation}\label{5.8}
\left(\frac{\partial \it{P_{eff}}}{\partial r_{+}}\right)_{x}\left(\frac{\partial \it{T_{eff}}}{\partial x}\right)_{r_{+}}-\left(\frac{\partial \it{P_{eff}}}{\partial x}\right)_{r_{+}}\left(\frac{\partial \it{T_{eff}}}{\partial r_{+}}\right)_{x}=0,
\end{equation}
substituting Eq.~(\ref{5.8}) into Eq.~(\ref{5.6}), we can get
\begin{equation}\label{5.9}
\left(\frac{\partial S}{\partial V}\right)_{\it{T_{eff}}}^{c}=\left(\frac{\partial S}{\partial V}\right)_{\it{P_{eff}}}^{c},
\end{equation}
substituting Eq.~(\ref{5.9}) into Eq.~(\ref{5.3}) to get universal Prigogine-Defay ratio $(\prod)$
\begin{equation}\label{5.10}
\prod=\left(\frac{\partial \it{P_{eff}}}{\partial \it{T_{eff}}}\right)_{S}/ \left(\frac{\partial \it{P_{eff}}}{\partial \it{T_{eff}}}\right)_{V}=\left(\frac{\partial S}{\partial V}\right)_{\it{P_{eff}}}/\left(\frac{\partial S}{\partial V}\right)_{\it{T_{eff}}}=1.
\end{equation}
Hence the phase transition occurring at $\it{T_{eff}}=\it{T_{eff}}^{c}$ is a second order equilibrium transition, similar to the AdS black hole \cite{Huaifan17,Banerjee004,Banerjee156,Niloofar Abbasvandi}. This is true in spite of the fact that the phase transition curves are smeared and divergent near the critical point.

\section{Conclusion and discussion}

Black hole physics, especially black hole thermodynamics, directly relates to various physical fields such as gravity, statistics, particles and field theory, which makes the subject attractive. It can be said that black hole physics has become the testing ground for various related theories. Black hole physics makes a profound and fundamental connection between gravity, statistics, particles, and quantum theory. Among them, black hole thermodynamics plays an important role. For de Sitter spacetime, since the radiation temperatures of the two horizons are generally different, spacetime is not in equilibrium, which generates difficulties to the study of the thermodynamic properties of de Sitter spacetime.

Recently, people have studied the thermodynamic properties of de Sitter spacetime under the assumption that the de Sitter spacetime entropy is the sum of the entropy of the black hole horizon and the cosmological horizon. Following the results, when the temperatures of the two horizons are the same, the effective temperature is generally not equal to the physical temperature of the two horizons. Therefore this result is unacceptable. In addition, treating the total entropy of RN-dS spacetime as the sum of the corresponding entropies of the two horizons is not proved theoretically.

The interaction between the black hole horizon and the cosmological horizon is considered to obtain the effective thermodynamic quantity of spacetime, and further analyze the thermodynamic properties of RN-dS spacetime. In Section 3, we find that RN-dS spacetime also has a continuous phase transition similar to that of Van de Waals system or  charged AdS black hole. For different RN-dS spacetime, as long as the position ratio of the two horizons of RN-dS spacetime is $x=0.656434$, the spacetime will undergo a continuous phase transition. The continuous phase transition point depends only on the position ratio of the two horizons, and is not a single function of the electric charge, mass, and cosmological constant of the black hole. When a continuous phase transition occurs, the electric potential at the black hole horizon is a constant, which is similar to AdS black holes~\cite{Xiong-Ying20,Xiong-Ying2019}. The phenomenon reflects that when RN-dS spacetime evolves to pure dS spacetime, the evolution process of spacetime is different due to the different charges in spacetime. Therefore, the electric potential at the black hole horizon plays a key role in the evolution of spacetime. Different RN-dS evolution processes are used to simulate the evolution process of the universe, and to open up a new way for seeking the internal factors that promote the accelerated expansion of the universe.
 When approaching the critical point, various thermodynamic functions either diverge or remain limited. Near the critical point, all thermodynamic functions can be written in the following form
\begin{equation}\label{6.1}
f(\epsilon)=A \epsilon^{\lambda}(1+B \epsilon^{y}+L),
\end{equation}
where $y>0$, the critical exponent of the function is defined as
\begin{equation}\label{6.2}
\lambda=\lim_{\epsilon\rightarrow 0}\frac{\ln f(\epsilon)}{\ln \epsilon}.
\end{equation}

At the critical point, the  heat capacity at constant pressure, the expansion coefficient and the isothermal compression coefficient, etc., all have a jump or infinite spike. One can use power functions to express the critical behaviors. Its power is called the critical exponent. Various ferromagnetic systems have common experimental laws in the vicinity of critical points.
In the fourth part, using the Landau theory of continuous phase transition, the critical phenomenon of RN-dS spacetime is discussed, and the order parameters introduced by similar ferromagnetic systems are found. For thermodynamic systems, different states of matter have different symmetry, phase transitions accompanied by changes in symmetry. On the other hand, symmetry is disorder. Phase transitions happen between ordered state and disordered state, or from higher symmetry to lower symmetry. The change is called symmetry breaking. The change in symmetry, or the change in order, can be a jump, which is a first-order phase change, or it can be a continuous change, which is a second-order phase change. The RN-dS spacetime reflects the order parameter of the degree of order of the system. From the other side, it is mapped out that the degree of order of microscopic particles in spacetime is related to the size of the electric potential where the particles are located. For ordinary particles, displacement polarization will occur under the action of an electric field. Harmonic polarization makes the arrangement of particles tend to be orderly. Under the action of electric potential, the microscopic particles in the spacetime of RN-dS also have displacement polarization and directional polarization similar to ordinary particles under the action of electric field. One issue needs further in-depth exploration.
This work reveals from one side that the spacetime microstructure of RN-dS has a microstructure similar to that of a ferromagnetic system, and these conclusions provide some help for people to explore the spacetime microstructure of RN-dS. The study of the microstructure of black holes is quantum. The foundation of the theory of gravity is a bridge for people to understand the relationship between quantum mechanics and the theory of gravity. In particular, the deep study of the microstructure of black holes will help to understand the basic properties of black hole gravity, and it is also very valuable for the establishment of quantum gravity.

\section*{Acknowledgments}

We would like to thank Prof. Zong-Hong Zhu and Meng-Sen Ma for the indispensable discussions and comments. This work was supported by the National Natural Science Foundation of China (Grant No. 12075143), supported by Program for the Natural Science Foundation of Shanxi Province, China(Grant No.201901D111315) and the Natural Science Foundation for Young Scientists of Shanxi Province,China (Grant No.201901D211441), supported by the Scientific Innovation Foundation of the Higher Education Institutions of Shanxi Province (Grant Nos. 2020L0471 and 2020L0472), and the Science Technology Plan Project of Datong City, China (Grant Nos. 2020153 and 2020155).

\end{document}